\documentstyle[epsf, twocolumn]{esapub}

\begin{document}
\setlength{\parindent}{0pt}
\setlength{\parskip}{ 10pt plus 1pt minus 1pt}
\setlength{\hoffset}{-1.5truecm}
\setlength{\textwidth}{ 17.1truecm }
\setlength{\columnsep}{1truecm }
\setlength{\columnseprule}{0pt}
\setlength{\headheight}{12pt}
\setlength{\headsep}{20pt}
\pagestyle{esapubheadings}
\newfont{\cm}{cmss10 scaled 1000}
\newfont{\cms}{cmss12 scaled 1000}
\newfont{\cmss}{cmssi12 scaled 1000}
\newfont{\cmsss}{cmss17 scaled 1000}


\title{FAR-IR AND MM PROPERTIES OF QUASARS}

\vspace{10mm}
\author{{\bf P. Andreani$^1$, A. Franceschini$^1$ G.L. Granato$^2$}
\vspace{2mm} \\ 
$^1$Dipartimento di Astronomia di Padova, vicolo dell'Osservatorio 5,
I-35142 Padova, Italy. \\ e-mail: andreani/franceschini@astrpd.pd.astro.it.\\
$^2$ Osservatorio Astronomico, vicolo dell'Osservatorio 5,
I-35142 Padova, Italy. \\
e-mail: granato@astrpd.pd.astro.it.}

\maketitle

\vspace{5mm} 
\begin{abstract} 

We use photometric data, from the optical to the {\it mm}, for a large sample of 
optically selected radio-quiet quasars, at low and high redshifts, 
to test emission models from circum-nuclear dusty torii around them.
Model parameters, such as dust mass, temperature distributions and
torus sizes are inferred, under the assumption that the dust is 
heated only by the central nuclear source.
Dependences of best-fit parameters on luminosity and redshift are 
studied and the contribution of dust in the host galaxy to the observed fluxes
is briefly mentioned.
Only tentative conclusions can be drawn from this yet large dataset and
detailed modelling,
because crucial spectral information is missing at wavelengths where the 
QSO dust emission is more prominent, i.e. at $\lambda \simeq 50$ to
600 $\mu m$.  Poor information in this wavelength range affects
our understanding not only of the high-redshift objects, but also
of the low-redshift sample.
We then expect fundamental contributions by the FIRST mission in this field, 
which is obviously related to the more general one of galaxy formation. 

\vspace{5mm}
Keywords: AGNs, dust emission, dust mass evolution
\end{abstract} 

\vspace{5mm}
\section{INTRODUCTION}

While disk galaxies appear to form their stellar content at moderate rates
during most of the Hubble time, there are no clear indications yet that
the expected bright phase of star formation in spheroidal galaxies has
been detected by optical searches. 
A possible explanation is that galaxies in this phase are
shrouded in dust, produced in large amounts by the early generations of
massive stars  (e.g. Franceschini et al. 1994; De Zotti et al. 1996).
\hfill\break
This idea of an early and quick metal and dust production 
affecting optical emission 
at least in bulge-dominated massive galaxies may receive some support by 
preliminary evaluations of the dust content in high-z quasars and of 
(very high) metal abundances in absorption line systems
"associated" to the QSO. These have shown 
that the close environments of high-z (both
radio-loud and radio-quiet) QSOs were the site of quick and
efficient star formation and metal production (Haehnelt \& Rees 1993;
Franceschini and Gratton 1997). 
Indeed, the presence of an AGN could even mask
the presence of a forming spheroid and complicate its
investigation, but can also be exploited to identify target areas where
to look for forming structures (galaxies, galaxy clusters and groups). 

Sub-{\it mm} - {\it mm} observations are then starting to shed light
into these phases.
The chance of detection of high-z objects at these wavelengths
is favoured by the strong and positive K-correction implied by the steeply
rising sub-{\it mm} spectra.
\hfill\break
Many high-z sources, most of which associated with active nuclei, 
have been recently detected in the millimeter (both in the continuum and CO
line emission; see Andreani et al. 1993; Chini and Kr\"ugel 1994; Dunlop
et al., 1994; Isaak et al. 1994; McMahon et al., 1994; Ivison 1995; Ohta
et al., 1996; Omont et al., 1996a and 1996b). With the exclusion of
radio-loud quasars, such radiation has been attributed to dust since the
observed spectrum, $F \propto \nu ^{3-4}$, is far too steep for any
dominant synchrotron self-absorbed contribution. 

In this contribution we make use of a large sample of optically selected
quasars at low and high redshifts, observed in the far-IR to {\it mm} region, 
to study dust emission in radio-quiet
objects and its possible dependence on redshift.
The basic assumption in this analysis is that the dust distribution, assumed
to have simple axial symmetry, either confined to the quasar nucleus 
or covering a sizeable fraction of the hosting galaxy, is illuminated
by the central nuclear source. Obvious evidence in support of this is
the dominance of the point-like nuclear source in the optical in both
high- and low-redshift objects.

This effort marks a significant improvement over previous studies, as it 
exploits the whole spectral information from the optical to the {\it mm},
the optical providing in particular an essential datum on the intensity 
of the radiation field illuminating the dust.

\vspace{5mm} 
\section{THE SAMPLE}

The sample consists of optically selected quasars with sub-{\it mm} and
{\it mm} observations. Far-IR data are from the IRAS PSC, and for most of
the sources co-added survey data were provided by IPAC based on the
SCANPI procedure.
{\it Mm} observations are partly collected from the literature
and partly taken by the authors with the IRAM 30m at Pico Veleta. 
The low-redshift sample consists mainly of PG radio-quiet quasars.
\begin{figure*}[!ht]
  \begin{center}
    \leavevmode
\epsfxsize=15.cm
\epsfysize=15.cm
\epsffile{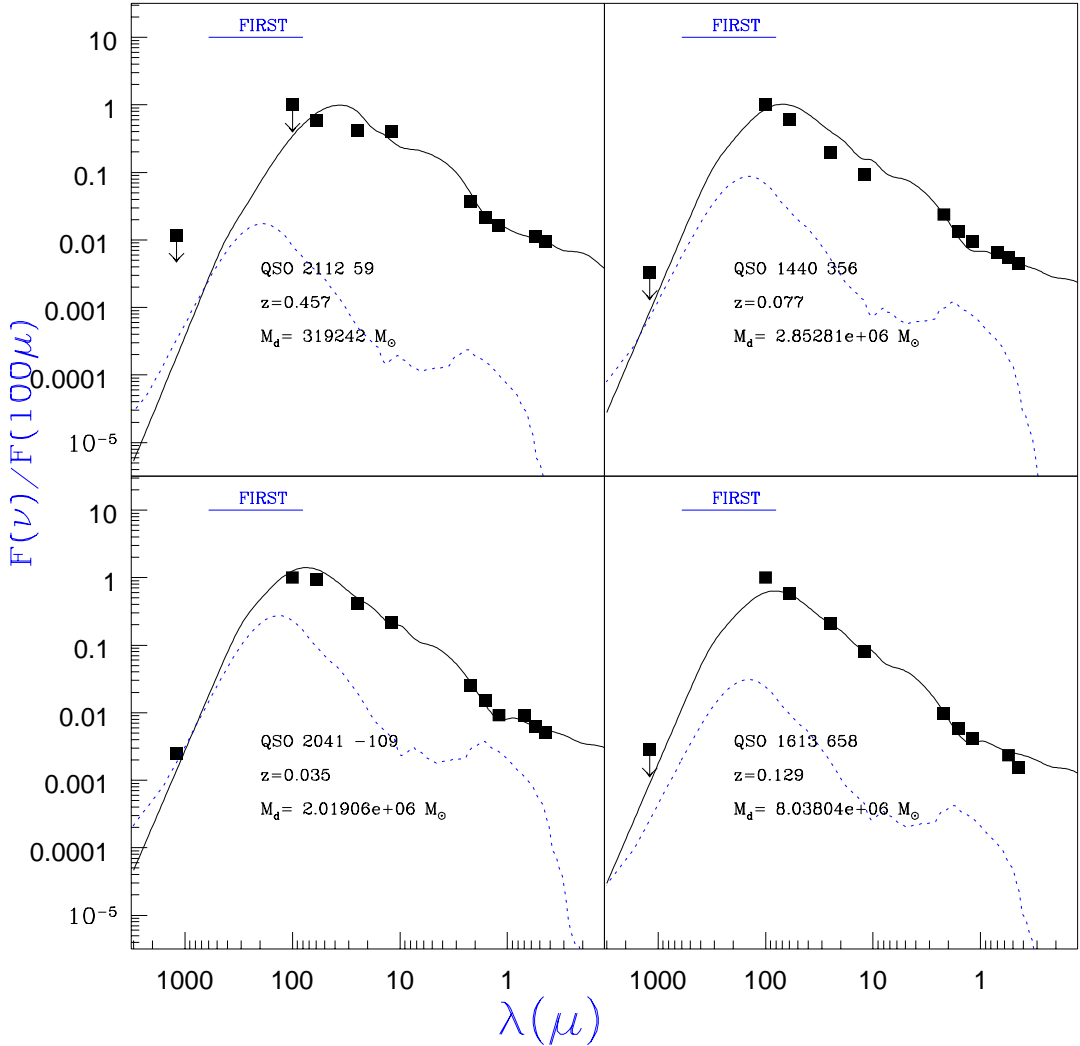}
  \end{center}
\caption{\em Observed SEDs of low-z AGNs fitted with the model described in
\S 3 (solid line) with the parameters ${r_{max} \over r_{in}}$, $\Theta$, $A_V$:
(a) 200, $0^\circ$, 18 (b)1000, $0^\circ$, 40 (c) 1500, $0^\circ$, 40
(d) 1500, $0^\circ$, 40. Dotted-line curve corresponds to the spectrum of
a typical spiral galaxy with a dust mass of $\sim 10^7 M_\odot$}
\end{figure*}

\vspace{10mm}
\section{MODELS OF DUST EMISSION IN QUASARS}

The model spectral energy distributions (SEDs) used to fit the observational 
data were computed with the numerical code described by Granato
\& Danese (1994), which solves the radiative transfer equation in a 
circumnuclear dust distribution.
This step is required since in the torii predicted by unified models the
dust emission is self--absorbed in the near-- and mid--IR.  The
code is quite flexible about the geometry and composition of the
dusty medium, the only restriction being axial symmetry, and thus allows
a wide exploration of the parameter space.

We assume in our exploration that the optical 
properties of dust are the same for all sources, with  
a standard Galactic composition.
The inner radius $r_{in}$ of the dust distribution is set by the grain
sublimation condition, at $T_s=1500$ for graphite and $T_s=1000$ for
silicates. This translates into the condition 
$r_{in}\sim 0.5 \sqrt{L_{46}}$ pc, where
$L_{46}$ is the luminosity of the primary optical--UV emission in units
of $10^{46}$ erg $s^{-1}$.
\hfill\break
The details of the model, as well as the effect of the various free
parameters, have been widely discussed by Granato \& Danese 1994 and in
Granato, Danese \& Franceschini (1997). Here we focus on
the simplest geometry which is in reasonable agreement with the available
observations, the ``flared disc'', in which the scale height of the
torus along the z-axis increases linearly with the radial distance, and
the dust distribution within the disc is homogeneous. Model parameters
are then {\it (i)} the outer radius $r_{max}$ of the dust
distribution; {\it (ii)} the absorption $A_V$ along typical obscured
directions; and {\it (iii)} the covering factor $f=\cos(\Theta_h)$ of 
the torus, where $\Theta_h$ is the half opening angle of the dust--free cones.
In practice, the free parameters in the fit were the former two, the 
last one having been fixed to $f=0.8$, a value not inconsistent with 
requirements of unified models of AGNs (Granato et al. 1997).
\begin{figure*}[!ht]
  \begin{center}
    \leavevmode
\epsfxsize=15.cm
\epsfysize=15.cm
\epsffile{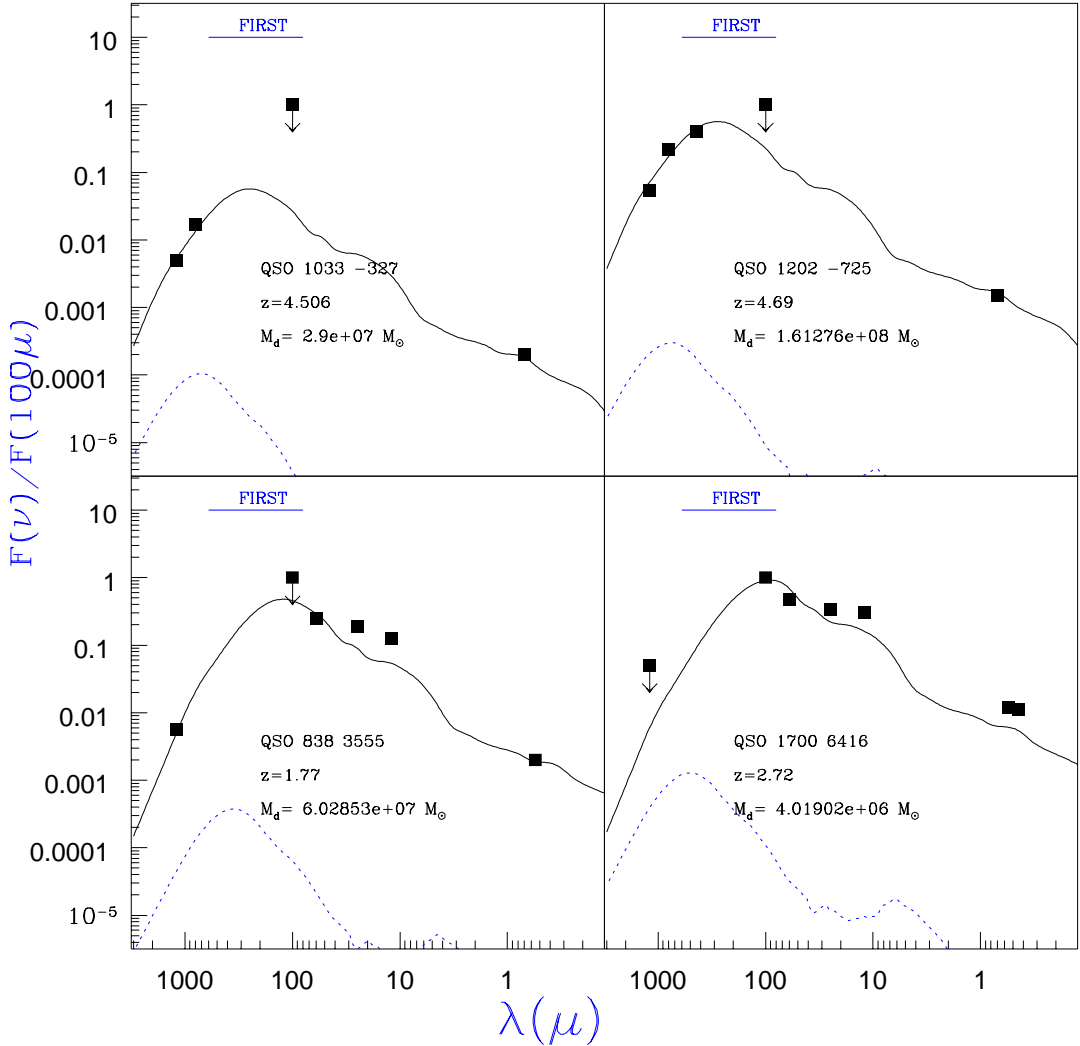}
  \end{center}
\caption{\em As figure 1 but for high-z AGNs 
(a) 600, $0^\circ$, 40 (b) 800, $0^\circ$, 40 (c) 600, $0^\circ$, 30 (d) 200,
$0^\circ$, 18 }
\end{figure*}

\begin{figure*}[!ht]
  \begin{center}
    \leavevmode
\epsfxsize=14.cm
\epsfysize=22.cm
\epsffile{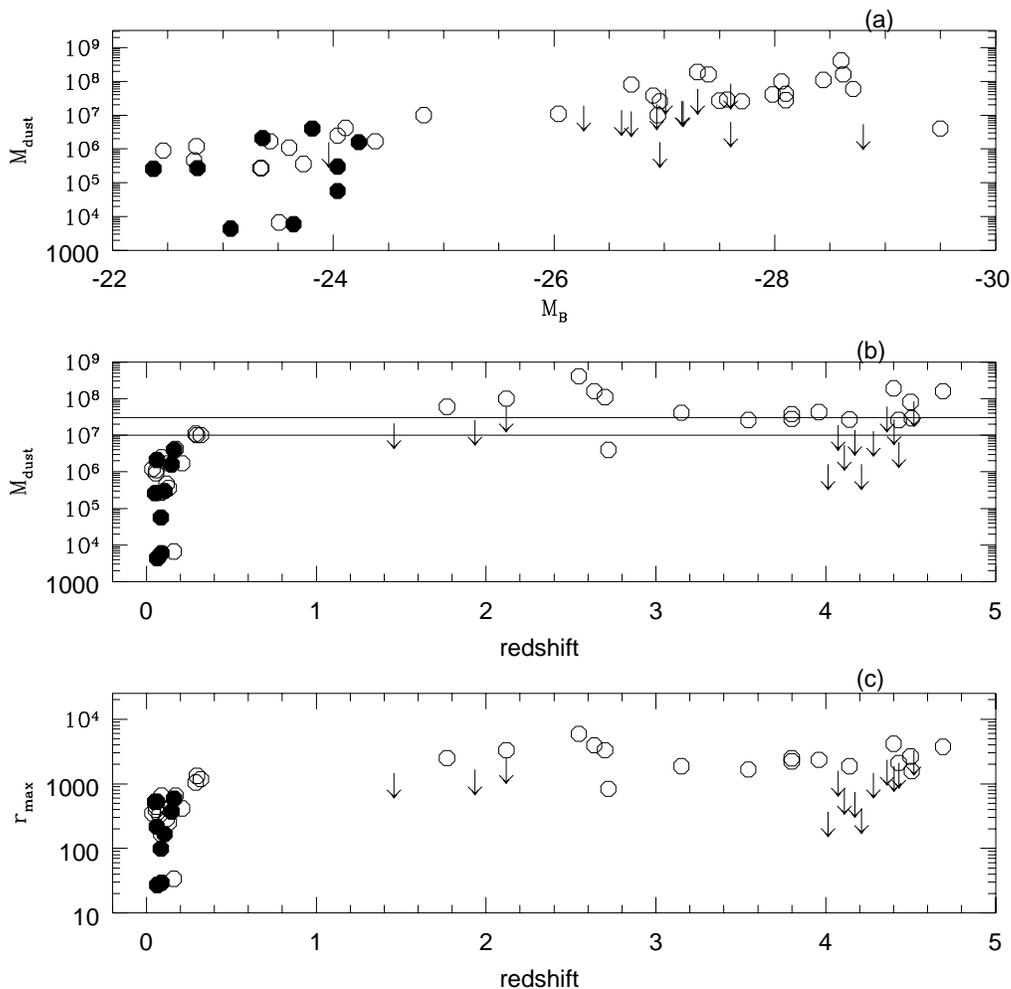}
  \end{center}
\caption{\em Dust masses versus (a) B-band absolute magnitude and (b)
redshift. (c) Maximum size of the dust emission region as a function of
redshift. Filled circles correspond to objects for which the presence of
a host galaxy is required to improve the fit at $\lambda > 60 \mu m $.
Horizontal lines in (b) show the range of dust masses found in nearby
spirals.} 
\end{figure*}

Roughly speaking, the first parameter is related to the broadness of the
IR bump arising from dust reprocessing, whilst the second controls
mainly the near-IR slope of the SEDs as observed from obscured
directions as well as its anisotropy.  Therefore $A_V$ is in principle
constrained by the SEDs of obscured AGNs or by testing the anisotropy of
mid--IR emission. The covering factor is obviously related to the
relative power reprocessed by the dust in the IR bump
and the primary optical continuum flux.
\hfill\break
In this simple geometry the total dust mass is given by the product
between the disc volume $V=\frac{4}{3} \pi  \cos \Theta_h$ and the
constant dust density $\rho=\frac{\tau_\nu}{k_\nu (r_{max}-r_{in})}$,
where $k_\nu$ is the absorption coefficient per unit dust mass
and $\tau _\nu$ the equatorial optical thickness.
This gives:

\begin{equation}
\frac{M_{\rm dust}}{M_\odot} \sim 0.2 \, ({r_{max} \over r_{in} })^2
({L \over L_{46}})^{3/2} \cos\Theta_h \, A_V
\end{equation}

Unless otherwise stated, all quantities are computed assuming  
H$_0$=50 km/s/Mpc and a flat universe ($q_0=0.5$).

\vspace{10mm}
\section{RESULTS}

Figures 1 and 2 show typical spectra of 4 low-z and 4 high-z sources.
Some information on the source 
and model parameters are reported in the figure panels and captions.
The thick solid curves are the best fits SEDs, while the dotted line is 
the broad-band optical-{\it mm} spectrum of a typical spiral galaxy, used for 
comparison. Clearly, while for local objects the hosting galaxy can 
contribute to some extent at the longest wavelengths, for high-z quasars 
the observed spectra are far in excess of what would be expected
from a redshifted SED of a normal galaxy. This reflects in part the
much increased power of the illuminating source, following the 
well-known evolution with $z$ of the primary continuum.
\hfill\break
The overall quality of the fits is remarkably good, if we consider 
that only two parameters ($r_{max}$ and $A_V$) are free to model the
spectral shape.

Figure 3a plots the best-fit dust mass, $M_{d}$, as a function of 
the B-band absolute magnitude, M$_B$, showing the expected trend
of increasing $M_d$ at increasing M$_B$. A similar trend is observed
with the AGN bolometric luminosity. 
\hfill\break
Figure 3b shows the evolution of the AGN M$_d$ versus redshift. 
The filled circles correspond to those objects for which the
contribution of the host galaxy is important. 
\hfill\break
Figure 3c plots the best-fit outer radius $r_{out}$ of the dust distribution 
versus redshift. $M_d$ and $r_{out}$ are very tightly related in our 
model, both being precisely constrained by the long wavelength ({\it mm} or 
sub-{\it mm}) flux data.
Therefore, the plots in Figs. 3b and 3c display quite a similar 
behaviour.
\hfill\break
In both cases the low-z objects show widely dispersed values 
in the parameters, with typically moderate values for the dust mass,
$M_d\sim 10^5-10^7\ M_\odot$ and from 10 pc to 1 kpc for $r_{max}$.
For the high-z subsample, typical values for $M_d$ and $r_{max}$ are 
larger, with $M_d$ typically in excess of $10^7$ and $r_{max}$ in excess 
of 1 kpc. 

Note that the procedure of the dust mass estimation is
quite a robust one: we believe that uncertainties in the model
parameters affect its value by less than a factor two, and the basic
conclusions are not affected. 

\vspace{5mm}
\section{DISCUSSION}

Some limitations of the present analysis have to be considered
before drawing any conclusions.

The first is that
the observational SEDs are interpreted assuming that most of the IR-{\it mm} 
flux comes from dust illuminated by the AGN itself and that the dust in 
the flared disc is homogeneously distributed from the inner sublimation 
surface up to kpc-scale distances.

For most of the local objects only upper limits
to the {\it mm} fluxes are available. So, it cannot be excluded that for some 
of these we are missing significant contributions 
of cold dust from the hosting galaxy at $\lambda \geq 100\mu m$ (
most of the {\it mm} observations were collected with large
antennas fed with single channel bolometers, having therefore a field of view
limited to a few arcsec).
Indeed, Danese et al. (1992) and Rowan-Robinson (1995) 
already suggested that the FIR spectrum at $\lambda \geq 60 \mu m$ in 
low-luminosity AGN may be contributed by cold dust in the surrounding galaxy 
situated at large radial distances from the AGN and with typical            
dust masses of 1 to 3 $\ 10^7$ M$_\odot$. 
\hfill\break
The relative contribution
at far-infrared wavelengths of the nuclear and extended component
will only be settled by high resolution mapping with planned large
{\it mm} arrays (e.g. LSA).

On the other hand, the SEDs of high-z object SEDs are less constrained, as
only a few spectral data are available for each source, due to the fact 
that IRAS was not sensitive enough to detect a significant number of them
(a couple of exceptions appear in Fig. 2). In these cases the
sub-{\it mm} data are the only information we have on the circum-nuclear
interstellar medium, IRAS providing a weak constraint on the dust
temperature distribution.
Comparison of our model to the data suggests the existence of a large 
gaseous structure
around the nucleus (with 1-5 kpc radius), including a dust mass of up
to a few 10$^8$ M$_\odot$, and a corresponding gas masses of typically 
$10^{10}$ to $10^{11}$ M$_\odot$. 
\hfill\break
We note, in particular, the large values of the outer radius of the dust 
distribution, which are dictated by the combined constraints set by the
{\it mm} flux, the primary optical field intensity and the IRAS upper limits.
Such extended distributions indicate that they are not merely 
circum-nuclear dust torii, as seen in local Seyfert 
galaxies, but have rather scale-lengths comparable with those of the
bulge of the putative host galaxy.


In principle, we would expect that our estimate of the dust mass $M_d$ in 
quasars and the size $r_{max}$ 
of the dust distribution should be unaffected by the 
power of the primary continuum and by its evolution with cosmic time.
In practice an obvious bias operates in Figure 3: the more luminous AGNs
found at the higher z are also likely to be hosted by higher mass 
galaxies.
Small mass objects at high-z escape detection and the lower right part of the
plot cannot be sampled by present instruments. 
\hfill\break
What is clear is that no high mass objects are found in the nearby
universe. 

A way to explain the trends observed in Figs. 3b and 3c would be to
assume that we are sampling the same class of objects.
Evolution in the dust mass content in quasars could then be related to the 
history of galaxy evolution, as revealed by optical deep surveys:
a {\it downsizing} process is taking place,
such that more massive galaxies form at higher
redshift, followed by a sequence of less and less massive objects forming
at lower and lower redshift down to the formation of dwarfs.

However, this opens the question of the fate of the dust: where has all
this dust gone? Was it swept away from the quasar? 
Was it consumed to form stars in a late phase of star formation?
Note that the difference between low and high redshift objects would be
further enhanced in an open universe: dust masses at redshifts larger
than 2 for $\Omega \sim 0.01$ are higher by a factor between 2 and 5.

The alternative possibility is that we are observing at high- and 
low-redshifts two distinct populations.
Environmental conditions at high redshifts could favour the formation 
of higher
mass black holes, while low-z AGNs may be related to the refueling of small
dead black holes in late type galaxies, where gas is still available
(Haehnelt \& Rees 1993).


Because of a selection bias (all sources were optically
selected), no edge-on objects were found in the present sample:
the nuclear source is not totally obscured from the circumnuclear torus.
Obscured sources, of which IRAS F10214 could represent the prototype,
can only be found from FIR/sub-{\it mm} surveys such as those expected
with the FIRST Satellite.
\hfill\break
An abundant population of dusty objects could heretofore escape
detection. Are many early quasars obscured by the products of massive
galaxy-forming starbursts?
Such objects might observationally appear very much like ultraluminous
infrared galaxies, so that sub-{\it mm} photometric surveys by FIRST
represent the only way to discover them. 

\section*{ACKNOWLEDGEMENTS}
We would like to thank L. Danese and S. Cristiani for instructive discussions.
Part of the data used in this work were taken with the SCANPI procedure,
developed by the NASA Archival center for IRAS Satellite (IPAC) operating
by JPL.

\end{document}